\documentstyle[multicol,aps,prl,floats]{revtex}

\def\L{\hat{\cal L}}
\begin{document}
\draft
\title{Geometric zero mode for the Kraichnan's problem of
passive scalar advection}
\author{E. Balkovsky}
\address{Physics Department, Weizmann Institute of Science,
Rehovot 76100, Israel }
\date{\today}
\maketitle
\begin{abstract}
We find explicitly a zero mode of the Kraichnan operator at two dimensions.

\end{abstract}
\begin{multicols}{2}

We consider the problem of passive scalar advection in multi-scale
$\delta$-correlated in time velocity field. As was shown in
\cite{68Kra-a}, the $n$-point
correlation function of a passive scalar satisfies the following closed
differential equation
\begin{equation}
-\L F_n=\Phi_n,\quad \mbox{where}\quad
\L=\sum_{i\not=j}{\cal K}_{\alpha\beta}(r_{ij})
\nabla_i^\alpha \nabla_j^\beta
\label{z1}\end{equation}
and $\Phi_n$ is the function which determines the pumping and can be  
expressed via the characteristics of the random force and
lower correlation functions of the passive scalar. 
A tensor ${\cal K}_{\alpha\beta}$ is given by
\begin{equation}
{\cal K}_{\alpha\beta}=Dr^{-\gamma}\left(\frac{d+1-\gamma}{2-\gamma}r^2
\delta_{\alpha\beta}-r_{\alpha}r_{\beta}\right).
\end{equation}
Here $\gamma$ is a parameter which can vary from zero to two.
It is believed, that 
deep inside convective interval the correlation function behaves as
$F_n\sim r^{\gamma n} (L/r)^{\Delta_n}$, where $r$ is a characteristic
distance between points and $\Delta_n$ is an anomalous exponent, which
makes deviation from normal scaling. As it was found out in 
\cite{95CFKLb,95GK}
this deviation arises due to zero modes of the operator $\L$, which are
solutions of the equation $\L Z=0$. That's why it is important to know
such functions. However, it is highly non-trivial problem to find them.
These modes were studied on the basis of perturbative
analyses in the cases of large $d$ \cite{95CFKLb} and $\gamma$ close to two
\cite{95GK}.
They were found explicitly in the non-scaling logarithmic case
$\gamma=0$ \cite{95BCKL}. 

Here we present a three-point zero mode which exists
at $d=2$ and arbitrary $\gamma$. 
The zero mode can be build as follows.
Let us consider the following object made of three points $r_1$, $r_2$,
and $r_3$ 
\begin{equation}
s=\epsilon_{\mu\nu}r_{12}^\mu r_{13}^\nu.
\end{equation}
Here $\epsilon$ is antisymmetric tensor and $\bbox{r}_{ij}=\bbox{r}_i-
\bbox{r}_j$.
It is trivially checked that $s$ is zero mode of the operator $\L$.
It has simple geometric meaning of the area of
the triangle. However, it changes sign under permutation of points and
therefore is zero after symmetrization. One can cure this by taking
absolute value of $s$ and introducing $Z(r_1,r_2,r_3)=|s|$, which is
automatically symmetric. Obviously, $Z$ satisfies the equation
$\L Z=0$ when either $s>0$ or $s<0$, having problems when $s=0$. 
First, it has jump in derivatives at $s=0$. 
Second, due to this jump it
may even not satisfy homogeneous equation, due to appearance of the term 
$A\delta(s)$ in the result of the action of $\L$ on $Z$. However, one can
easily calculate the coefficient $A$ and find that it is zero. As for
the angular singularity, it may be present in the problems of such a kind and
should be smeared by the diffusive term in the equations. In exactly
solvable case $\gamma=0$ it was found \cite{95BCKL} that zero mode entering
solution has singularity of the same type. 

Let us write the expression for $Z$ in variables $r_{12}$, $r_{13}$, and
$r_{23}$
\begin{equation}
Z=\sqrt{2r_{12}^2r_{13}^2+2r_{12}^2r_{23}^2+2r_{13}^2r_{23}^2-
r_{12}^4-r_{13}^4-r_{23}^4}.
\end{equation}
We don't know yet which importance this zero mode has for the theory.
It is not clear, for example if it enters the solution for the $n$-point
correlation
functions or not. One can note, however, that even if it does, as a
three-point object it disappears
from structural functions $\langle(\theta_1-\theta_2)^{2n}\rangle$
\cite{95GK,95CF}. 
Note that $Z$ scales as $r^2$, therefore it wins over normal scaling exponent
$n\gamma/2$ only for $\gamma>4/n$.

One may try to generalize this construction. First, at $d=2$ we can build a
four-point function $s_{1234}=\epsilon_{\mu\nu}r_{12}^{\mu}r_{34}^{\nu}$.
However, taking of absolute value makes singularity stronger, since
the manifold $s_{1234}=0$ is more complicated. As a result, coefficient
$A$ in front of $\delta$-function is not zero anymore for $\gamma\not=0$.
It makes this function not a true zero mode.  
Let us note special importance which areas $s_{ijkl}$
have at $\gamma=0$ \cite{95BCKL}: in corresponding variables
operator $\L$ does not contain differentiations over $s_{ijkl}$ for
any $n$, and becomes simple enough to solve the problem
exactly for the correlation function of any order.
For higher dimensions, one could try to construct volumes 
$v=\epsilon_{\mu_1\ldots\mu_d}r_{12}^{\mu}\cdots r_{ij}^{\mu_d}$
which are zero
modes of the operator. However, after we take absolute value, again 
$\delta$-function presents in case $\gamma\not=0$, since the manifold
$v=0$ has nontrivial structure. Thus, we see that zero modes
of such a kind exist only for two dimensional problem and can be built
only for three-point geometry.

\acknowledgments
I am grateful to G. Falkovich, V. Lebedev, M. Chertkov, I. Kolokolov, and
D. Gutman for helpful discussions.

\end{multicols}

\end{document}